\documentclass{article}

\usepackage{url}
\usepackage{amsmath}
\usepackage{amssymb}
\usepackage{amsfonts}
\usepackage{amsthm}
\usepackage[all]{xy}
\usepackage{verbatim}
\usepackage{graphicx}

\newcommand{\half}{\frac{1}{2}}

\newcommand{\sfr}[2]{^{#1}/_{#2}}

\title{For the Monomer-Dimer Problem on Triangular and Hexagonal Lattices, the New $p$-Expansion}
\author{Paul Federbush \thanks{Department of Mathematics, University of Michigan, Ann Arbor,
 MI 48109-1043, \emph{email}:pfed@umich.edu} }
\date{}

\begin{document}
\maketitle

\begin{abstract}
In a recent paper S. Friedland and the author presented a formal expression for $\lambda_d(p)$ of the monomer-dimer problem on a $d$-dimensional rectangular lattice, which involved a power series in $p$. Herein, we find simlar expressions for $\lambda_{\textrm{tri}}(p)$ and $\lambda_{\textrm{hex}}(p)$, the entropy per site for triangular and hexagonal lattices.
\end{abstract}

The current work grew out of a treatment of the dimer problem on a rectangular lattice in $d$-dimensions by the author: leading to a formal asymptotic expansion for $\lambda_d$
\begin{align}
\lambda_d \sim \half\ln(2d)-\half+\sum_{k=1}^\infty \frac{c_k}{d^k}
\end{align}
where the first three $c_i$ have been computed [1]. Working with S. Friedland this formalism was extended to treat the monomer-dimer problem (on the same lattices) obtaining the formal expansion
\begin{align}
\lambda_d(p)\sim \half(p\ln(2d)-p\ln p - 2(1-p)\ln(1-p) -p)+\sum_{k=1}^\infty \frac{c_k(p)}{d^k}
\end{align}
where the first three polynomials, $c_k(p)$, are known [4]. The authors came to believe it is better to organize (2) into a power series in $p$
\begin{align}
\lambda_d(p)\sim \half(p\ln(2d)-p\ln p-2(1-p)\ln(1-p)-p)+\sum_{k=2}^\infty a_k(d)p^k
\end{align}
Here the first six polynomials in $1/d, a_k(d)$ are known and given in [4]. In fact, we now believe the expression on the right side of (3) converges to $\lambda_d(p)$ for all physical values, $d=1,2,\cdots, 0\leq p \leq 1$! For $d=1$ the first six terms in (3) are known to be correct, for $d=2$ the terms we have agree very well with numerical studies, [4]. In [5] I have given a rigorous proof that the sum in (3) converges for small enough $p$. 

In this paper I present the analog of (3) for the two dimensional triangular and hexagonal lattices (see [2] and [3] for general background), and for comparison the two dimensional rectangular lattice. For any lattice the difficult work is computing the corresponding cluster expansion kernels $\bar{J}_i$ defined in [1], by machine computations increasingly lengthy as $i$ increases. Once these are known, the $p$ expansions are easy computations---short runs of short Maple programs. Equations (13), (15), and (16) of [5] are useful here. For the rectangular lattice the first six $\bar{J}_i$ were computed and given for all dimensions in [1]. $\bar{J}_7$ was recently computed (but only in two dimensions), [7]. For the triangular and hexagonal lattice the $\bar{J}_i$ are given in [6], for the triangular the first five, and for the hexagonal the first six. Recently we computed an additional term for each of these series, using the same algorithm highlighted in [7].  For each of the three lattices we present the results.

\section*{Rectangular Lattice}
The first six $\bar{J}_i$ have the values, in sequence: $0, \sfr{1}{16},\sfr{1}{48}, -\sfr{9}{512}, -^{23}/_{1280}, \sfr{25}{3072}, \sfr{299}{14336}$, from [1] and [7]. 
\begin{align}
\lambda_2(p)\sim& \half(p\ln(4)-p\ln p -2(1-p)\ln(1-p)-p) +\frac{4}{2}\cdot\left(\frac{1}{2\cdot 1}\left(\frac{p}{4} \right)^2\right.\\
&\left.+\frac{1}{3\cdot 2}\left(\frac{p}{4} \right)^3+\frac{7}{4\cdot 3}\left(\frac{p}{4} \right)^4+\frac{41}{5\cdot4}\left(\frac{p}{4} \right)^5+\frac{181}{6\cdot 5}\left(\frac{p}{4} \right)^6+\frac{757}{7\cdot 6}\left(\frac{p}{4} \right)^7\right)\nonumber
\end{align}

\section*{Triangular Lattice}
The first six $\bar{J}_i$ have the values, in sequence $0,\sfr{1}{24},0,-\sfr{31}{1728},-\sfr{13}{6480},\sfr{10}{729}$.
\begin{align}
\lambda_{\textrm{tri}}(p)\sim& \half(p\ln(6)-p\ln p-2(1-p)\ln(1-p)-p)\\
&+\frac{6}{2}\cdot\left(\frac{1}{2\cdot1}\left(\frac{p}{6}\right)^2-\frac{3}{3\cdot2}\left(\frac{p}{6}\right)^3-\frac{11}{4\cdot 3}\left(\frac{p}{6}\right)^4+\frac{1}{5\cdot 4}\left(\frac{p}{6}\right)^5+\frac{7\cdot13}{6\cdot 5}\left(\frac{p}{6}\right)^6\right)\nonumber
\end{align}

\section*{Hexagonal Lattice}
The first seven $\bar{J}_i$ have values in sequence: $0,\sfr{1}{12},\sfr{1}{27},-\sfr{7}{216},-\sfr{23}{405},-\sfr{5}{1458},\sfr{395}{5103}$.
\begin{align}
\lambda_{\textrm{hex}}(p)\sim& \half(p\ln (3) - p\ln p - 2(1-p)\ln(1-p)-p)+\frac{3}{2}\cdot\left(\frac{1}{2\cdot 1}\left(\frac{p}{3}\right)^2\right. \\
&\left.+\frac{1}{3\cdot2}\left(\frac{p}{3}\right)^3+\frac{1}{4\cdot 3}\left(\frac{p}{3}\right)^4+\frac{1}{5\cdot4}\left(\frac{p}{3}\right)^5+\frac{11}{6\cdot 5}\left(\frac{p}{3}\right)^6+\frac{5\cdot17}{7\cdot 6}\left(\frac{p}{3}\right)^7\right)\nonumber
\end{align}
The expressions (4),(5), (6) are arranged to exhibit some very interesting patterns, not all of which features we yet understand. We think a key thing to study is the relationship of $\lambda(p)$ for each of these lattices to $\lambda(p)$ for a regular tree graph with the same degree (a Bethe lattice). I think I will be able to explain the denominator factors as of 'simple', 'geometric' origin.  Can one determine the full $p$ dependence in any of these cases?

\end{document}